\documentclass[aps,prl,preprintnumbers,amsmath,amssymb,superscriptaddress,twocolumn]{revtex4-2}
\DeclareMathAlphabet{\mathpzc}{OT1}{pzc}{m}{sl}   
\usepackage[pdfborder={0 0 0}]{hyperref}
\usepackage{amsmath,amssymb,upgreek}
\usepackage{graphicx}
\usepackage{color,bm}
\usepackage{natbib}
\usepackage{float}
\usepackage{comment}
\usepackage{bbm,verbatim}
\usepackage{booktabs}
\usepackage{array}
\usepackage{appendix}
\usepackage{multirow}
\usepackage{caption}
\usepackage{setspace} 
\begin{document}

\definecolor{dkblue}{rgb}{0.0,0.55,0.88}
\definecolor{dkmagenta}{rgb}{0.8,0.3,0.8}
\newcommand{\av}[1]{\langle\mspace{1mu}{#1}\mspace{1mu}\rangle}
\newcommand{\bs}[1]{\boldsymbol{#1}}
\newcommand{\ba}[1]{\text{#1}}
\newcommand{\bl}[1]{\text{#1}}
\newcommand{\bss}[1]{\vec{\boldsymbol{#1}}}
\newcommand{\bt}{\vec{\boldsymbol{\uptau}}}

\newcommand{\bT}{\overset{\text{\tiny$\boldsymbol{\leftrightarrow}$}}{\boldsymbol{T}}}
\newcommand{\bI}{\overset{\text{\tiny$\,\boldsymbol{\leftrightarrow}$}}{\boldsymbol{I}}}

\newcommand{\bII}{{\stackrel{\text{\scriptsize$\boldsymbol{\leftrightarrow}$}}
                 {\boldsymbol{\mathbb{I}}}}}
\newcommand{\bTT}{{\stackrel{\text{\scriptsize$\boldsymbol{\leftrightarrow}$}}
                 {\boldsymbol{\mathbb{T}}}}}
\newcommand{\beps}{{\stackrel{\text{\tiny$\boldsymbol{\leftrightarrow}$}}
                 {\boldsymbol{\mathbb{\epsilon}}}}}
\newcommand{\doint}{\displaystyle\oint}

\newcommand{\bMM}[2]{{\stackrel{\text{\scriptsize$\boldsymbol{\leftrightarrow}$}}
                 {\boldsymbol{\mathbb{M}}}_{#1}^{\text{\raisebox{-1.8ex}{$#2$}}}}}
\newcommand{\bNN}[2]{{\stackrel{\text{\scriptsize$\boldsymbol{\leftrightarrow}$}}
                 {\boldsymbol{\mathbb{N}}}_{#1}^{\text{\raisebox{-1.8ex}{$#2$}}}}}
\newcommand{\bLL}[2]{{\stackrel{\text{\scriptsize$\boldsymbol{\leftrightarrow}$}}
                 {\boldsymbol{\mathbb{L}}}_{#1}^{\text{\raisebox{-1.8ex}{$#2$}}}}}
\newcommand{\bOO}[2]{{\stackrel{\text{\scriptsize$\boldsymbol{\leftrightarrow}$}}
                 {\boldsymbol{\mathbb{O}}}_{#1}^{\text{\raisebox{-1.8ex}{$#2$}}}}}
\newcommand{\bU}[1]{{\stackrel{\text{\scriptsize$\boldsymbol{\leftrightarrow}$}}
                 {\boldsymbol{\mathbb{U}}}^{\text{\raisebox{-1.8ex}{$#1$}}}}}

\newcommand{\be}{\,\vec{\boldsymbol{e}}}
\newcommand{\brp}{\vec{\boldsymbol{r}}^{\,\prime}}
\newcommand{\bh}{\,\vec{\boldsymbol{h}}}
\newcommand{\ft}[1]{\bs{\mathcal{#1}}}
\newcommand{\td}{\,\mbox{\rm d}}
\newcommand{\eg}{\,\mbox{\it e.g.,\;}}
\newcommand{\dint}{\displaystyle\int}
\newcommand{\dsum}{\displaystyle\sum}
\newcommand{\ddint}{\overset{\mspace{15mu}\infty}{\underset{\mspace{-20mu}-\infty}{\dint\!\!\!\dint}}}
\newcommand{\ddintl}[1]{\underset{\mspace{0mu}{#1}}{\dint\!\!\!\dint}}
\newcommand{\pd}[2]{\frac{\partial{#1}}{\partial{#2}}}
\newcommand{\pdi}[1]{\partial_{#1}}
\newcommand{\then}{\;\;\;\Longrightarrow\;\;\;}
\renewcommand{\-}{\text{\,\rule[0.55ex]{5mm}{.5pt}}\,}
\renewcommand{\.}{\boldsymbol{\cdot}}
\newcommand{\x}{\boldsymbol{\times}}
\newcommand{\wz}{\widetilde{z}}
\newcommand{\tc}[1]{{\;\stackrel{(#1)}{\text{\raisebox{-0.5ex}{$\boldsymbol{\cdot\cdot}$}}}\;}}
\newcommand{\tcj}[1]{{\,\stackrel{(#1)}{\text{\raisebox{-0.5ex}{$\boldsymbol{:}$}}}\,}}
\newcommand{\tcc}[1]{{\,%
                     \text{\raisebox{-1.0ex}{$\stackrel{(#1)}{\stackrel{\text{\small\raisebox{-0.3ex}{$\boldsymbol{:}$}}}{\times}}$}}%
                     \;}}

\newcommand{\bM}{{\bs M}}
\newcommand{\bN}{{\bs N}}
\newcommand{\bX}{{\bs X}}
\newcommand{\br}{{\bs r}}
\newcommand{\etheta}{\,{\bs e}_{\theta}}
\newcommand{\ephi}{\,{\bs e}_{\phi}}
\newcommand{\er}{\,{\bs e}_{r}}
\newcommand{\opL}{\boldsymbol{\hat{L}}\,}
\newcommand{\buu}[1]{{\stackrel{\text{\tiny$\boldsymbol{\leftrightarrow}$}}
                 {\boldsymbol{u}}^{\text{\raisebox{-0.8ex}{$#1$}}}}}
\newcommand{\myi}{\mathpzc{i}\mspace{-4.2mu}{\text{\raisebox{-0.0ex}{$\imath$}}}\mspace{-5.5mu}{\text{\raisebox{0.8ex}{$\boldsymbol{\cdot}$}}}}

\newcommand{\revzhx}[1]{\textcolor{black}{#1}}
\newcommand{\revyx}[1]{\textcolor{black}{#1}}
\newcommand{\revyxx}[1]{\textcolor{black}{#1}}
\newcommand{\revzhxx}[1]{\textcolor{black}{#1}}
\newcommand{\revlin}[1]{\textcolor{black}{#1}}
\newcommand{\revlina}[1]{\textcolor{black}{#1}}
\newcommand{\revy}[1]{\textcolor{black}{#1}}
\newcommand{\revchen}[1]{\textcolor{black}{#1}}

\title{Universal \revlin{parity and \revlina{duality} asymmetries-based} optical force/torque framework} 


\author{Xu Yuan}
\thanks{These authors contributed equally to this work.}
\affiliation{School of Electronic Engineering, Guangxi University of Science and Technology, Liuzhou, Guangxi 545006, China}

\author{Xiaoshu Zhao}
\thanks{These authors contributed equally to this work.}
\affiliation{State Key Laboratory of Surface Physics and Department of Physics, Fudan University, Shanghai 200433, China}

\author{Jiquan Wen}
\affiliation{School of Electronic Engineering, Guangxi University of Science and Technology, Liuzhou, Guangxi 545006, China}

\author{Hongxia Zheng}
\email[]{hxzheng18@fudan.edu.cn}
\affiliation{School of Electronic Engineering, Guangxi University of Science and Technology, Liuzhou, Guangxi 545006, China}
\affiliation{State Key Laboratory of Surface Physics and Department of Physics, Fudan University, Shanghai 200433, China}
\affiliation{Guangxi Key Laboratory of Multidimensional Information Fusion for Intelligent Vehicles, Liuzhou, Guangxi 545006, China}

\author{Xiao Li}
\email[]{lixiao@ust.hk}
\affiliation{Department of Physics, Southern University of Science and Technology, Shenzhen 518055, China}
\affiliation{Department of Physics, The Hong Kong University of Science and Technology, Hong Kong, China}

\author{Huajin Chen}
\email[]{huajinchen13@fudan.edu.cn}
\affiliation{School of Electronic Engineering, Guangxi University of Science and Technology, Liuzhou, Guangxi 545006, China}
\affiliation{State Key Laboratory of Surface Physics and Department of Physics, Fudan University, Shanghai 200433, China}
\affiliation{Guangxi Key Laboratory of Multidimensional Information Fusion for Intelligent Vehicles, Liuzhou, Guangxi 545006, China}

\author{Jack Ng}
\affiliation{Department of Physics, Southern University of Science and Technology, Shenzhen 518055, China}

\author{Zhifang Lin}
\affiliation{State Key Laboratory of Surface Physics and Department of Physics, Fudan University, Shanghai 200433, China}
\affiliation{Collaborative Innovation Center of Advanced Microstructures, Nanjing University, Nanjing 210093, China}



\begin{abstract}
Understanding how the structured incident light interacts with the inherent properties of the manipulated particle 
\revlin{and governs} the optical force/torque exerted \revlin{is} 
a cornerstone in the \revchen{design of optical manipulation techniques}%
\revlin{, apart from its theoretical significance}. Based on the Cartesian multipole expansion theory, 
we establish a framework for optical force/torque \revlin{exerted on an arbitrary sized bi-isotropic (chiral) spherical    
particle immersed in generic monochromatic optical fields. Rigorous expressions are thus derived} which 
explicitly bridges such mechanical effects of light with particle-property-dependent coefficients and \revlin{``force/torque source" quantities} 
that \revlin{characterize the} 
incident light structures. 
\revlin{Such quantities, totalled only 12, are 
quadratic in terms of electric and magnetic field vectors, 
among which are linear and angular momenta, gradient of energy density, spin density, and helicity.} 
\revlin{
They are further organized into four categories based on 
their parity (P) and duality (D) \revchen{symmetries} \revlina{
and shown to couple with a particle with 
different P and D \revchen{symmetries} to induce optical force/torque.}} 
This classification specifies the \revlin{symmetry-breaking criteria required} 
to induce optical force/torque, offering a 
\revlin{promising} roadmap for engineering the optical manipulation. 
\end{abstract}

\maketitle
\revzhx{Optical manipulation, which entails the precise control of mechanical effects induced 
by the electromagnetic field through optical forces/torques, has become an indispensable tool across diverse research fields 
\cite{gao2017optical,grier2003revolution,zhuang2024physics}. The cornerstone of tailoring these forces/torques lies in a profound 
comprehension of the intricate interplay \revyxx{between the light structure and particle characteristics.} 
Significant milestones have been made within the dipole regime\revlin{. The radiation pressure, nascent knowledge} 
of the mechanical effect of light, is inherently linked to the \revzhxx{energy} flow} (\revlin{also termed} time-averaged Poynting vector) 
\cite{PMID:13072497,loudon2012contributions}. 
\revlin{The} well-known optical tweezers find their origins in 
\revlin{the inhomogeneity of} light intensity \cite{ashkin1986observation}. 
Meanwhile, the generation of torque has primarily been traced to optical spin 
\cite{beth1936mechanical,ahn2018optically,rashid2018precession,reimann2018ghz,jin20216}. 
These insights, \revzhxx{together with the force \revlin{attributed to} 
Belinfante's spin momentum \cite{albaladejo2009scattering,liberal2013near,bliokh2014extraordinary,bekshaev2015transverse,nori2016nphys,xnyu2023}, 
``stored energy" flow \cite{xu2019azimuthal,bliokh2014extraordinary,bekshaev2015transverse}, 
spin density \cite{wang2014lateral,hayat2015lateral}, and local optical helicity density inhomogeneity
\cite{canaguier2013mechanical,jin2024harnessing} et. al.,} are \revlin{mostly facilitated by the 
expressions of optical force/torques 
derived within the dipole approximation 
\cite{bliokh2014magnetoelectric,golat2024optical}}.
\setlength{\parskip}{3pt} 

\revlin{Although providing a lucid physical picture for 
the mechanical effects in light-matter \revchen{interactions}
and thus extensively applied to trace the physical \revchen{origins,} 
the dipolar framework of optical force/torque falls short 
when high-order multipoles are excited in a Mie particle with size beyond the dipole regime, 
thus obscuring some intriguing phenomena such as optical pulling \cite{chen2011optical}. 
It was until recently that, 
based on the multipole expansion theory,
some further understanding of optical force and torque are developed in some specific light fields 
beyond the dipole approximation \cite{zhou2022observation,zheng2021optical,xnyu2023,xu2024gradient},
contributing several important pieces to the puzzle in the quest for a universal and complete physical picture of the mechanical effects 
in general.
Nevertheless, the full picture for generic structured light fields and manipulated \revchen{particles} of arbitrary size and composition
remains lingeringly elusive, due to the inherent mathematical complexity.}

In this paper, \revlin{we establish a
comprehensive} theoretical framework for optical force and torque, based fundamentally on the \revchen{parity (P) and duality (D) symmetries} of 
both the particles and the incident light structures.
\revlin{The framework enables an in-depth} understanding of the mechanical effects of light. \revlin{It is} general, universal for 
\revlin{bi-isotropic (chiral) spherical} particles of arbitrary size and composition, 
\revlin{illuminated by} 
arbitrary monochromatic optical fields. 
We consider a reciprocal bi-isotropic particle that does not break \revchen{time-reversal} symmetry. 
Its constitutive relation is described by \cite{bohren2008absorption}, 
\revlin{
\begin{equation}\label{co-re}
\begin{split}
\bs D=\varepsilon_0\varepsilon\,\bs E+i\kappa\sqrt{\varepsilon_0\mu_0}\,\bs H, \\[.5ex] 
\bs B=\mu_0\mu\bs\, H-i\kappa\sqrt{\varepsilon_0\mu_0}\,\bs E,
\end{split}
\end{equation}
where $\varepsilon_0$ ($\mu_0$) and $\varepsilon$ ($\mu$) represent the vacuum and relative permittivity (permeability), respectively,
and $\kappa$ describes the particle chirality. It is noted that $\kappa\ne0$ denotes the \revchen{P} symmetry-breaking of the particle.}

\begin{widetext}
	
\begin{figure}[H]
\centering
\includegraphics[width=\textwidth]{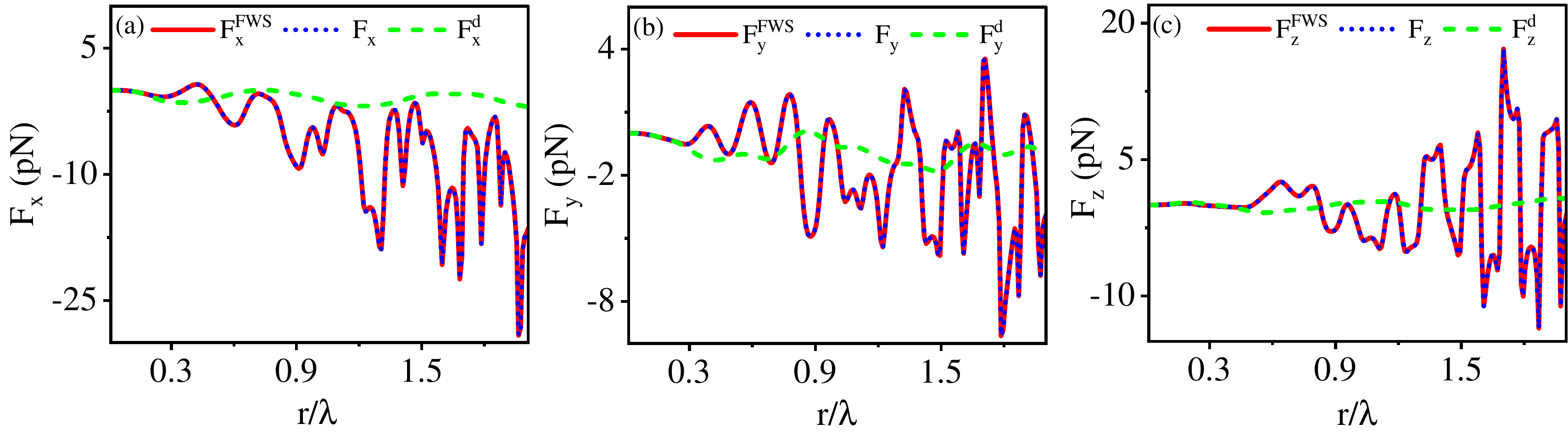}
\caption{\revlin{The three Cartesian components of the optical force exerted on \revlin{a spherical chiral particle} as a function of
particle radius $r$. The particle, with $\kappa=0.3$, \revchen{$\varepsilon=2.53$,} and $\mu=1$, 
resides in \revy{water} and is illuminated by the interference fields composed of five plane waves. 
The five plane waves, all with the same wavelength $\lambda=1.064~\, \mu$m and electric field amplitude $E_0=7.53 \times 10^5$ V/m, 
are randomly generated with different propagation \revchen{directions} and polarizations. 
The $F_j$ (blue dotted lines), with $j=\text{x, y, z}$, are calculated by Eqs.~(\ref{FX})-(\ref{FXir}). They exhibit
perfect agreement with $F_j^\text{FWS}$ (red solid lines) computed by a full-wave simulation (FWS) method 
using the generalized Lorenz-Mie theory \cite{jack2005,gouesbet2011generalized} and the Maxwell stress tensor \cite{zangwill2013modern,jackson2021classical}.
Also shown in the figures (green dashed lines) are $F_j^\text{d}$ evaluated with the dipole approximation 
\revlina{that sets $l_{\max}=1$ in Eq.~(\ref{FXir})}. 
Significant discrepancy between $F_j^\text{d}$ and $F_j^\text{FWS}$ 
is seen when \revlina{$r\gtrsim 0.3\lambda$.}}}
\label{fig-force}
\end{figure}\vspace{10pt}
	
\end{widetext}
\revlin{Based on the classical electromagnetic theory},
the time-averaged optical force exerted on an illuminated particle can be \revyxx{formulated} as 
	\begin{equation}\label{FX}
		\av{\bs F}=\displaystyle\sum_{\text{X}}\av{\bs F^\text{X}_\text{int}}+\displaystyle\sum_{\text{X}}\av{\bs F^\text{X}_\text{rec}},
	\end{equation}
where $\text{X}=(\text{0}),\,(\text{da}),\,(\text{pa}), \, \text{and} \, (\text{pda})$ denotes four different types of force, 
each \revlin{associated with a distinct P and D \revchen{asymmetries} on light field and particle (see below).} 
\revlin{The first sum in Eq.~(\ref{FX})} 
is the interception force,  which can be physically
\revlin{understood as resulting from} the process that light intercepted by the particle will transfer its momentum to the particle, 
thus exerting an interception force on it. \revlin{The second sum 
represents the recoil force}, originating  from light is re-emitted \revyxx{by various multipoles induced 
		on the illuminated particle} \revlin{\cite{nieto2010optical,chen2011optical,jiang2015universal,jiang2016decomposition}.} 
\revlin{They can be worked out to be}
\begin{widetext}
	\begin{subequations}\label{FXir}
		\begin{align}
			\av{\bs F^\text{X}_\text{int}}&=\revlin{\sum_{l=1}^{l_{\max}}}
			\Biggr\{\text{Re}\left[\alpha_l^\text{X}\right]\left({k}^{2}+\dfrac{\Delta}2\right)^{l-1}\nabla W^\text{X} 
                 +2\,\omega\,\text{Im}\left[\alpha_l^\text{X}\right]\left({k}^{2}+\dfrac{\Delta}2\right)^{l-1}\bs{P}^\text{X}\Biggr\}, \label{FXira}\\[2mm]
			\av{\bs F^\text{X}_\text{rec}}&=\revlin{\sum_{l=1}^{l_{\max}}}
			\Biggr\{\mathcal{I}_{\,l}^\text{X}\left({k}^{2}+\dfrac{\Delta}2\right)^{l-1}\nabla W^\text{X}+
             2\,\omega\,\mathcal{T}_{\,l} ^\text{X}\left({k}^{2}+\dfrac{\Delta}2\right)^{l-1}\bs{P}^\text{X}
			+k\,\omega\,\mathcal{L}_{\,l}^{\text{X}}\left({k}^{2}+\dfrac{\Delta}2\right)^{l-1}\bs{S}^\text{X}\Biggr\}, \label{FXirb}
		\end{align}
	\end{subequations}
\end{widetext}\vspace{10pt}
\revlin{where $k$ is the wavenumber in the background, $l_{\max}$ represents the highest multipole orders required to achieve convergence, and
$\omega$ denotes the angular frequency. The sum runs over all orders of the electric and magnetic multipoles}.
The role of the manipulated particle are described by the coefficients 
$\alpha_l^\text{X}$, $\mathcal{I}_{\,l}^\text{X}$, $\mathcal{T}_{\,l}^\text{X}$, and 
$\mathcal{L}_{\,l}^\text{X}$, which are functions of the Mie coefficients \cite{bohren2008absorption} $a_l$, $b_l$, and $c_l$
\revlina{(with $l=1,2,\cdots,l_{\max}$)}.
\revlin{The interception force} $\av{\bs F^\text{X}_{\text{int}}}$ arises from the interaction 
between the oscillating multipoles and the incident optical field, thereby being linked to \revlin{the ``force source" quantities 
$\nabla W^\text{X}$ and $\bs{P}^\text{X}$ }
through a \revlin{generalized polarizability $\alpha^\text{X}_{\,l}$} linearly dependent on the Mie coefficients.
The recoil force $\av{\bs F^\text{X}_{\text{rec}}}$, \revlin{on the other hand,}
originates from the interplay between oscillating multipoles excited on the particle, 
demonstrating a \revlin{quadratic} dependence on 
Mie coefficients in its coefficients $\mathcal{I}^\text{X}_{\,l}$, $\mathcal{T}^\text{X}_{\,l}$, and $\mathcal{L}^\text{X}_{\,l}$
\revlin{associated, respectively, with the ``force source" quantities $\nabla W^\text{X}$, $\bs{P}^\text{X}$, and, also,
$\bs{S}^\text{X}$.  All these 12 ``force source" quantities, which show no dependence on the particle's property,
characterize the light structure \cite{golat2024}. 
They all exhibit quadratic dependence on the electric ($E$) and magnetic ($H$) fields and actually exhaust all possible vector forms quadratic in
$E$ and $H$-fields, which suffice to determine the optical force on an isotropic particle universally.} 
\revlin{It is also noted that while the energy density-like inhomogeneity $\nabla W^\text{X}$ and the momentum density-like
$\bs{P}^\text{X}$ generate 
the mechanical effect of light through both the interception force and recoil force, 
the spin density-like $\bs{S}^\text{X}$ only \revchen{couples} with particles during the recoil process to induce a recoil force.}
\revlin{Finally, $\Delta$ in Eq.~(\ref{FXir}) represents the Laplacian operator, with the
the exponent denoting the times that the operator \revchen{acts} on the three types of \revchen{``force source"} quantities.}

\revlin{To validate Eqs.~(\ref{FX}) and (\ref{FXir}), we have performed extensive
calculation of optical force on a chiral particle versus particle radius, with the particle immersed in
interference fields formed by a large variety of plane waves randomly generated.
Perfect agreement is found between the results by \revchen{Eqs.~(\ref{FX}) and (\ref{FXir})}
and those by a full-wave simulation (FWS) method that combines
the generalized \revchen{Lorenz-Mie theory with the Maxwell stress tensor method} 
\cite{jack2005,gouesbet2011generalized,zangwill2013modern,jackson2021classical}.
Typical results are shown in Fig.~\ref{fig-force},
highlighting the robust reliability of 
Eqs.~(\ref{FX}) and (\ref{FXir}) for generic structured light \cite{forbes2021} and particles of generic size, and, also,
demonstrating the deficiency in the dipole approximation when particle radius $r>0.3\lambda$, with $\lambda$ being the manipulating wavelength.
The power of Eqs.~(\ref{FX}) and (\ref{FXir}) is thus corroborated for universally
tracing the intricate physics of the optical force profoundly into the light structures and particle characteristics.}
\revlina{Now we turn to the four different categories of ``force source'' quantities and
analyze from the perspective of P and D \revchen{(a)symmetries}. 
It is emphasized that the analysis below is performed for isotropic particle of arbitrary size and composition.
The 
$\text{X}=(\text{0})$ category exhibit fundamental field characteristics. They are given by 
\begin{subequations} \label{X=(0)terms}
	\begin{align}
		W ^{(0)}
		&=\dfrac14{\varepsilon_0}\big(\bs{E}\.\bs{E}^*+Z_0^2 
                                \bs{H}\.\bs{H}^*\big),\\[1.ex]
		\bs{P} ^{(0)}		&=\dfrac{\varepsilon_0}{4\,\omega}\text{Im}\left[\left(\nabla\bs{E}\right)\.\bs{E}^*
                     +Z_0^2\left(\nabla\bs{H}\right)\.\bs{H}^*\right],\\[1.ex]
		\bs{S} ^{(0)}
		&=\dfrac{1}{4\,\omega\,c}\text{Re}\big[\bs{E}\times\bs{H}^*+\bs{E}^*\times\bs{H}\big],
	\end{align}
\end{subequations}
with $Z_0^2=\varepsilon_0/\mu_0$. They}
describe, respectively the \revzhx{time-averaged} energy densities,  canonical (or orbital) momentum density, 
and energy current density (Poynting vector).  
\revlina{Any particle, irrespective of its P or D symmetry characteristics, can perceive \revchen{the} category of ``force source'',
manifesting itself as a fundamental optical force.
The coefficients $\alpha_l^\text{(0)}$, $\mathcal{I}_l^\text{(0)}$, $\mathcal{T}_l^\text{(0)}$, and $\mathcal{L}_l^\text{(0)}$ in Eq.~(\ref{FXir})
are nonvanishing except for the cases with a perfect index and impedance matching to the background 
(e.g., $\varepsilon=1,\,\mu=1$, and $\kappa=0$ if the particle resides in vacuum).
The radiation pressure and the gradient force in conventional optical tweezers fall into this category.
On the other hand, 
the rest three categories of ``force source'', corresponding to $\text{X}=(\text{da}),\, (\text{pa}),\, \text{and}\, (\text{pda})$ and
singularizing some particular light \revchen{structures}, require the manipulated particle to possess 
certain P \revchen{or/and} D asymmetry to induce an optical force. }

\revlina{The ``force source'' quantities for $\text{X}=(\text{da})$ are
\begin{subequations}  \label{X=(da)terms}
	\begin{align}
		W ^\text{(da)}
		&=\dfrac14{\varepsilon_0}\big(\bs{E}\.\bs{E}^*-Z_0^2
               \bs{H}\.\bs{H}^*\big),\\[1.ex]
		\bs{P} ^\text{(da)}	
	&=\dfrac{\varepsilon_0}{4\,\omega}\text{Im}\left[\left(\nabla\bs{E}\right)\.\bs{E}^*-
                 Z_0^2\left(\nabla\bs{H}\right)\.\bs{H}^*\right], 
                 \\[1.ex]
		\bs{S} ^\text{(da)}
		&=\dfrac{1}{2\,\omega\,c}\text{Im}\left[\bs{E}\times\bs{H}^*-\bs{E}^*\times\bs{H}\right],
	\end{align}
\end{subequations} 
where $W^\text{(da)}$ ($\bs{P}^\text{(da)}$) represents the difference between the electric and magnetic parts of the 
energy density (canonical momentum density), in contrast to  
the sums of the corresponding electric and magnetic parts in Eq.~(\ref{X=(0)terms}).
The spin density-like $\bs{S}^\text{(da)}$ is actually the imaginary part of the complex Poynting vector. It delineates} 
the alternating flow of the so-called ``stored energy" (or reactive power) \cite{jackson2021classical}. 
\revlina{The quantities in Eq.~(\ref{X=(da)terms}) contribute to the optical force 
exclusively when the particle exhibits D asymmetry, as seen from their associated coefficients 
$\alpha_l^\text{(da)}$, $\mathcal{I}_{\,l}^\text{(da)}$, $\mathcal{T}_{\,l}^\text{(da)}$, and $\mathcal{L}_{\,l}^\text{(da)}$ 
that depends solely on the Mie coefficients.
To be specific, the ``force source'' quantities of the $\text{X}=(\text{da})$ category play a role in the optical force
only if $\varepsilon/\varepsilon_b\ne\mu/\mu_b$ \cite{chen2020lateral}, with $\varepsilon_b$ ($\mu_b$) 
denoting the relative permittivity (permeability) of the background.}

\revlina{The third type of the ``force source'' quantities, with $\text{X}=(\text{pa})$, consists of
\begin{subequations}
	\begin{align}
		W^\ba{(pa)}
		&=\dfrac{k}{2\,\omega}\text{Im}\left[\bs{E}\.\bs{H}^*\right],\\[1.ex]
		\bs{P}^\ba{(pa)}
		&=\dfrac1{4\,\omega\,c}\text{Re}\left[(\nabla \bs{H})\. \bs{E}^*-(\nabla \bs{E}^*)\.\bs{H}\right],\\[1.ex]
		\bs{S}^\ba{(pa)}
		&=\dfrac{i\,\varepsilon_0}{4\,\omega}\big(\bs{E}^*\times\bs{E}
                   +
                   Z_0^2\bs{H}^*\times\bs{H}\big),
	\end{align}
\end{subequations} 
with} $W ^\ba{(pa)}$, $\bs{P} ^\ba{(pa)}$, and $\bs{S} ^\ba{(pa)}$ being, respectively, the time-averaged helicity density, 
the real part of the complex chiral momentum densities \cite{bliokh2014magnetoelectric,golat2024optical}, 
and the spin angular momentum densities. 
\revlina{The quantity $\bs{P}^\ba{(pa)}$ is also written} \revyxx{as
$\bs{P}^\ba{(pa)}=k\left(\bs{S}_\text{e}+\bs{S}_\text{m}\right)-\dfrac{1}{2\,\omega\,c}\nabla\times\bs{P}$ 
in \cite{bliokh2014magnetoelectric,golat2024optical}, 
where \revlina{$\bs{S}_\text{e}=-\dfrac{i\,\varepsilon_0}{4\,\omega}\left(\bs{E}^*\times\bs{E}\right)$ and 
$\bs{S}_\text{m}=-\dfrac{i\,\mu_0}{4\,\omega}\left(\bs{H}^*\times\bs{H}\right)$} represent 
the electric and magnetic components of the spin angular momentum of the light field, and 
$\nabla\times\bs{P}=\dfrac{1}{2}\nabla\times\text{Re}\left[\bs{E}\times\bs{H}^*\right]$ 
represents the vortex of the light field.}
\revlina{These ``force source'' quantities 
couple with particle to generate the optical force only if
the particle exhibits P-symmetry breaking, i.e., for particle with $\kappa \neq 0$.}
\revlina{Intriguingly, 
the force associated with these quantities 
is a pure chiral force, namely, it reverses its direction for particles of opposite chirality.} 
\revlina{This characteristic 
is elucidated by fact that all} coefficients $\alpha^\text{(pa)}_{\,l}$, $\mathcal{I}^\text{(pa)}_{\,l}$, $\mathcal{T}^\text{(pa)}_{\,l}$, 
and $\mathcal{L}^\text{(pa)}_{\,l}$ in Eq.~(\ref{FXir}) \revlina{exhibit a linear dependence on the chiral Mie coefficient  $c_l$, 
which satisfies $c_l(-\kappa)=-c_l(\kappa)$ for any $l$.}
\revlina{This feature of this type of force} 
can, therefore, be utilized to separate particles with opposite handedness, 
which generally exhibit remarkably different biological activities, 
spanning pharmacokinetics to biotoxicology \cite{nguyen2006chiral}.\setlength{\parskip}{3pt} 

Finally, 
\revlina{the last type of ``force source'' quantities,} 
with $\text{X}=(\text{pda})$, are identified as
\allowdisplaybreaks
\revlina{
\begin{subequations}
	\begin{align}
W ^\ba{(pda)}
		&=\dfrac{k_0}{2{\,\omega}}\text{Re}\left[\bs{E}\.\bs{H}^*\right],\\[1.ex]
\bs{P} ^\ba{(pda)}
		&=\dfrac{1}{4\omega\,c}\text{Im}\left[(\nabla \bs{H})\. \bs{E}^*-(\nabla \bs{E}^*)\.\bs{H}\right],\displaybreak[0]\\[1.ex]
\bs{S} ^\ba{(pda)}		&=\dfrac{i\,\varepsilon_0}{4\,\omega}\big(\bs{E}^*\times\bs{E}
                           -Z_0^2\bs{H}^*\times\bs{H}\big), 
	\end{align}
\end{subequations}
where}
$W^\ba{(pda)}$, $\bs{P}^\ba{(pda)}$, and $\bs{S}^\ba{(pda)}$ characterize, respectively,  the time-averaged magnetoelectric energy density
\cite{bliokh2014magnetoelectric}, imaginary part of the complex chiral momentum density \cite{golat2024optical} 
(magnetoelectric momentum densities \cite{bliokh2014magnetoelectric}), 
and the difference between the electric and magnetic spin angular momentum density. 
In contrast to types \revchen{$\text{X}=(\text{pa})$ and $\text{X}=(\text{da})$,} \revlina{where the ``force source'' quantities couples exclusively to
particles with P and D \revchen{asymmetries}, respectively, to generate the optical force, the current type of ``force source'' quantities
requires that both P and D symmetries be broken on the particle to produce the force.} 
\revlina{In addition, this type of quantities manifest themselves exclusively in the recoil force, since
the coefficient $\alpha^\text{(pda)}_{\,l}=0$ in Eq.~(\ref{FXira}) turns out to be identically zero.}
This indicates that, \revlina{to induce this type of force,
multiple electromagnetic multipoles} 
must be excited on the particle. \revlina{Only the multipole interactions
between electric and magnetic multipoles of the same order 
as well as those of adjacent 
orders of the same type produce the optical force.} 
\revlina{The quantity 
$\bs{P}^\ba{(pda)}$, which was also written 
as $\bs{P}^\ba{(pda)}=\dfrac{1}{4\omega\,c}\nabla\times\text{Im}\left[\bs{E}\times\bs{H}^*\right]$, 
engender the optical force through the interplay between
multiploles of same type and with adjacent orders. As a result,  
within the dipole approximation where $l_{\max}=1$ is set in Eq.~(\ref{FXir}), 
$\bs{P}^\ba{(pda)}$ does not bring about any optical force, except for the nonreciprocal particle that possesses time-reversal asymmetry
\cite{bliokh2014magnetoelectric,golat2024optical}.}

\revlina{For the case beyond the dipole approximation,
it is obviously inadequate to expect that the optical force exerted on a large particle 
keeps depending exclusively on the local values of 
the ``force source'' quantities at the particle center.}
\revlina{By lengthy algebra, we manage to attribute the ``non-local'' effect to 
the operators 
\begin{equation} \label{operatorA}
\hat{\mathcal{A}}_l\equiv\Big({k}^{2}+\dfrac{\Delta}2\Big)^{l}.
\end{equation} 
When applied to the ``force source'' quantities, such operators encompass the effect of higher order multipoles while,
for their isotropy and P-symmetry, keeping
the symmetry features of all the ``force source'' quantities:} the energy or energy-like gradient $\nabla W^\text{X}$, 
the momentum or momentum-like $\bs{P}^\text{X}$, and spin or spin-like $\bs{S}^\text{X}$. 
\revlina{In addition, based on the potential theory \cite{Plessis1970AnIT}, we have,
for any function $f(\bs r)$ of the position vector $\bs r$,}
\revlina{
	\begin{equation*}
		\hat{\mathcal{A}}_lf(\bs r)\Big|_{\bs r=0}=\sum_{j=0}^{l}\dfrac{2j+1}{2^j}\,C^{\,l}_j\,
		k^{2(l-j)}\,\dfrac{\td^{2j}M_r(f)}{\td\mspace{1mu}r^{2j}}\bigg|_{r=0},
	\end{equation*}
where $C^{\,l}_j$ are the binomial coefficients,      
and $M_r(f)$ denotes the mean of the function $f$ 
over a spherical surface $S(r)$ of radius $r$ centered at $0$.} 
\revlina{In this sense, \revchen{Eq.~(\ref{FXir}) keeps} its physical elegance in expressing the optical force
in terms of the 12 ``force source'' quantities that fall into 4 categories based on P and D \revchen{symmetries}, 
while taking into account the contribution from the ``non-local effect''
due to the higer order multipoles beyond dipoles.}


The \revzhx{theoretical framework of} optical torque can \revzhx{similarly} be organized \revzhx{as}
	\begin{equation} \label{TX}
		\av{\bs T}=\displaystyle\sum_{\text{X}}\av{\bs T^\text{X}_\text{int}}+\displaystyle\sum_{\text{X}}\av{\bs T^\text{X}_\text{rec}},
	\end{equation}
\revzhx{also with $\text{X}=(\text{0}),\,(\text{da}),\,(\text{pa}),\,\text{and}\,(\text{pda})$ 
denoting four different types of torque with corresponding P or D symmetry-broken requirements.} 
The interception and recoil torques are respectively given by
\begin{widetext}
	\begin{subequations}\label{TXir}
		\begin{align}
			\av{\bs T^\text{X}_\text{int}}&=\sum_{l=1}^{l_\text{max}}
			\Biggr\{{a} ^\text{X}_{\,l}\left({k}^{2}+\dfrac{\Delta}2\right)^{l-1}\nabla \mathcal{W}^\text{X}+\omega\,{b} ^\text{X}_{\,l}\left({k}^{2}+\dfrac{\Delta}2\right)^{l-1}\bs{\mathcal{P}}^\text{X}+k\,\omega\,{d}^{\text{X}}_{\,l}\left({k}^{2}+\dfrac{\Delta}2\right)^{l-1}\bs{\mathcal{S}}^\text{X}\Biggr\},\\[2mm]
			\av{\bs T^\text{X}_\text{rec}}&=\sum_{l=1}^{l_\text{max}}
			\Biggr\{\mathcal{A} ^\text{X}_{\,l}\left({k}^{2}+\dfrac{\Delta}2\right)^{l-1}\nabla \mathcal{W}^\text{X}+\omega\,\mathcal{B} ^\text{X}_{\,l}\left({k}^{2}+\dfrac{\Delta}2\right)^{l-1}\bs{\mathcal{P}}^\text{X}+k\,\omega\,\mathcal{D}^{\text{X}}_{\,l}\left({k}^{2}+\dfrac{\Delta}2\right)^{l-1}\bs{\mathcal{S}}^\text{X}\Biggr\}.
		\end{align}
\end{subequations} 

\begin{table}[H]
\centering
\caption{The P and D \revchen{symmetries} broken requirements for, respectively, the manipulated particle and the impinging field, 
in the generation of optical force and torque on chiral particles of generic size and composition 
immersed in an arbitrary monochromatic \revlina{structured} light field. 
The complete light structures \revlina{are characterized by 12 ``force/torque source" quantities, which fall into four categories
according to their P and D \revchen{symmetries}. 
The effects beyond dipole approximation are taken into account by introducing the operators $\hat{\mathcal{A}}_l$ 
given in Eq.~(\ref{operatorA}), which
keep the symmetry features of the ``force/torque source'' quantities.}
The symbol ``$+$" (``$-$") represents the even (odd) behavior of 
\revlina{the ``force/torque source'' quantities} under the \revchen{P} inversion or \revchen{D} transformation.
The fundamental $\text{X}=(\text{0})$ type force/torque can be induced on \revlina{any} 
particle \revlina{except for the case the particle exhibits a perfect index and impedance matching to the background.
This is denoted by ``No'' in the table, meaning that the optical \revchen{force/torque} can be induced with no need of}
either $\kappa \neq 0$ or $\varepsilon/\varepsilon_\text{b} \neq \mu/\mu_\text{b}$.
The generation of optical force/torque \revlina{of} 
types $\text{X}=(\text{da}),\,(\text{pa}),\,\text{and}\,(\text{pda})$ requires 
the corresponding P or D symmetry \revlina{breaking}, 
as denoted by ``Yes". 
\revlina{The \revchen{entry} marked by ``$\#$" \revchen{means} that the corresponding ``force/torque source'' quantities are irrelevant to the optical torque.
It is noted that a field intensity gradient will induce  a torque on a P symmetry breaking (also known as chiral or bi-isotropic) particle \cite{wen2024optical}.}}
\includegraphics[width=\textwidth]{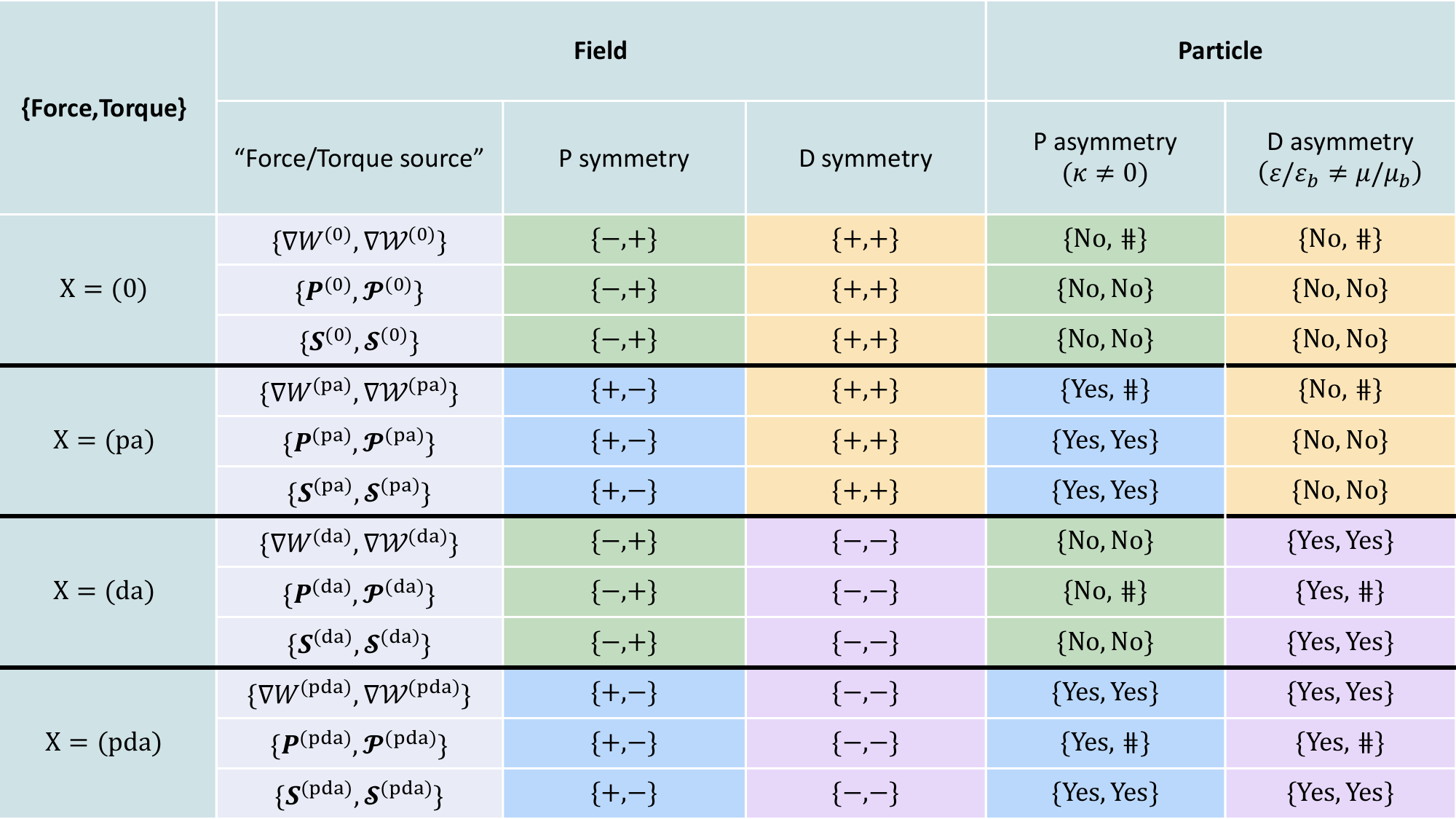}
\label{mytable}
\end{table}


\revzhx{For clarity, we adopt scripted letters $\mathcal{W} ^\text{X}$, $\bs{\mathcal{P}}^\text{X}$, and $\bs{\mathcal{S}}^\text{X}$ to denote field quantities in different torque \revchen{types}, instead of regular Latin letters $\nabla W^\text{X}$, $\bs{P}^\text{X}$, and $\bs{S}^\text{X}$ for force as presented in Eq.~(\ref{FXir}) to avoid confusion. 
 It is important to highlight that the \revy{\revchen{``torque source"} quantities} responsible for the four \revchen{torque types do not align with ``force source" ones, but they follow the relationship}}
	\begin{equation}
	\begin{array}{llll}
		\mathcal{W} ^{(0)}=W ^\ba{(pa)}, & \qquad
		\mathcal{W} ^\text{(da)}=\ W ^\ba{(pda)},&\qquad 
		\mathcal{W} ^\text{(pa)}=W ^{(0)},&\qquad
		\mathcal{W} ^\text{(pda)}=W ^\text{(da)}, \\[2mm]
		\bs{\mathcal{P}} ^{(0)}=\bs{P} ^\ba{(pa)},&\qquad
		\bs{\mathcal{P}} ^\text{(da)}=\bs{P} ^\ba{(pda)},&\qquad
		\bs{\mathcal{P}} ^\text{(pa)}=\bs{P} ^{(0)},&\qquad
		\bs{\mathcal{P}} ^\text{(pda)}=\bs{P} ^\text{(da)},\\[2mm]
		\bs{\mathcal{S}} ^{(0)}=\bs{S} ^\ba{(pa)},&\qquad
		\bs{\mathcal{S}} ^\text{(da)}=\bs{S} ^\ba{(pda)}, &\qquad
		\bs{\mathcal{S}} ^\text{(pa)}=\bs{S} ^{(0)},&\qquad
		\bs{\mathcal{S}} ^\text{(pda)}=\bs{S} ^\text{(da)}.
	\end{array}
\end{equation}
\end{widetext}
\revchen{However, some certain light structures, including those associated with ``torque source" quantities $\mathcal{W} ^\text{(0)}$, $\mathcal{W} ^\text{(pa)}$, $\bs{\mathcal{P}}^\text{(da)}$, and $\bs{\mathcal{P}}^\text{(pda)}$ that contribute to the optical force, do not give rise to optical torque. This is because} the interactions between multipoles of the same electric (magnetic) type with adjacent orders, which play crucial roles in optical force mechanism, \revchen{fail to produce} optical torque. 
Hence, the helicity density gradient associated light structures, pivotal in chiral sorting via optical force \cite{canaguier2013mechanical}, are unable to generate optical torque \revyxx{on bi-isotropic chiral particles}, let alone be anticipated to aid in chiral sorting through torque mechanisms. 
Furthermore, the surviving \revchen{$\mathcal{W} ^\text{(da)}$, $\mathcal{W} ^\text{(pda)}$, $\bs{\mathcal{P}}^\text{(0)}$, and $\bs{\mathcal{P}}^\text{(pa)}$,} only contribute to optical torque beyond dipole regime with $l\geq2$, as analyzed by the coefficients ${a} ^\text{X}_{\,l}$, ${b} ^\text{X}_{\,l}$, $\mathcal{A} ^\text{X}_{\,l}$, and $\mathcal{B} ^\text{X}_{\,l}$ in Eq. (\ref{TXir}). This makes their potential for optical torque tailoring historically overlooked for a long time until recently \cite{xu2024gradient},  
in contrast to the widespread recognition of the spin-like quantities $\bs{\mathcal{S}}^\text{X}$ \cite{ahn2018optically,rashid2018precession,reimann2018ghz,jin20216,chen2017optical,canaguier2013mechanical}.
\revzhx{In addition, it is intriguing to observe that \revchen{the light structures, associated with ``force source" quantities $\nabla W^\text{X}$, $\bs{P}^\text{X}$, and $\bs{S}^\text{X}$ with $\text{X}=(\text{pa}),\,(\text{pda})$, is conducive to chiral sorting from the perspective of force, however, which} now cannot separate opposite chirality with optical torque, manifested as \revchen{``torque source"} quantities $\nabla \mathcal{W} ^\text{(X)}$, $\bs{\mathcal{P}}^\text{X}$, and $\bs{\mathcal{S}}^\text{X}$ with $\text{X}=(\text{0}),\,(\text{da})$. }
Conversely, \revchen{``force source" quantities for the cases $\text{X}=(\text{0})$ and $(\text{da})$}, unfavorable for chiral separation based on optical force, can actually aid in differentiating chirality through optical torque. This implies that by constructing an optical field with \revchen{strong light structures for X=(0), (da) or X=(pa), (pda) type,}  one can consistently employ it for chirality discrimination and identification, utilizing either optical force or torque.
\revzhx{Lastly, the \revchen{``force/torque source" quantities for $\text{X}=\text{(pda)}$} type \revchen{demonstrates its mechanical effects, including both optical force and torque, exclusively} via the recoil physical process, as indicated by coefficients $\alpha^\text{(pda)}_{\,l}=0$ in Eq.~(\ref{FXir}a), alongside ${a}^\text{(pda)}_{\,l}=0$, ${b}^\text{(pda)}_{\,l}=0$, and ${d}^\text{(pda)}_{\,l}=0$ in Eq. (\ref{TXir}a).
}

\revzhx{The symmetry requirements of particles for both optical force and torque for four types encompassing  $\text{X}=(\text{0}),\,(\text{da}),\,(\text{pa}),\,\text{and}\,(\text{pda})$, as previously analyzed, notably characterized by the particle properties $\kappa \neq 0$ for P symmetry broken and $\varepsilon/\varepsilon_\text{b} \neq \mu/\mu_\text{b}$ for D symmetry broken, are consolidated in Table \ref{mytable}. This table also encapsulates the \revchen{P and D symmetries} characteristics of the \revy{``force/torque source" quantities}.} The electric and magnetic fields $\bs{E}$ and $\bs{H}$ under the parity inversion ($\hat{P}$) and the duality transformation ($\hat{D}$) are given by \cite{bliokh2014magnetoelectric}
	\begin{equation}
		\{{\hat{P}}, \hat{D}\}
		\begin{pmatrix}
			\sqrt{\varepsilon}\bs{E} \\
			\sqrt{\mu}\bs{H}
		\end{pmatrix}=
		\begin{Bmatrix}
			\begin{pmatrix}
				-\sqrt{\varepsilon}\bs{E} \\
				\sqrt{\mu}\bs{H}
			\end{pmatrix} ,
			\begin{pmatrix}
				\sqrt{\mu}\bs{H} \\
				-\sqrt{\varepsilon}\bs{E}
			\end{pmatrix} \\
		\end{Bmatrix}.
\end{equation} 
\revy{It is noteworthy that the ``force/torque source'' quantities requires the action of the operators $\hat{\mathcal{A}}_l$ given in Eq.~(\ref{operatorA}) to account for effects beyond the dipole approximation, \revchen{meanwhile maintaining its P and D symmetries. This is because} the operator is conserved under both P and D transformations.}
\revzhx{Quantities that exhibit symmetry or antisymmetry, equivalently even or odd behavior, under parity inversion and duality transformation \cite{calkin1965invariance,fernandez2013electromagnetic},  are denoted by ``$+$" and ``$-$", respectively. The symmetry properties of \{$\nabla W^\text{X}$, $\bs{P}^\text{X}$, $\bs{S}^\text{X}$\}  or \{$\nabla \mathcal{W}^\text{X}$, $\bs{\mathcal{P}}^\text{X}$, $\bs{\mathcal{S}}^\text{X}$\} for each class $\text{X}=(\text{0}),\,(\text{da}),\,(\text{pa}),\,\text{and}\,(\text{pda})$ always remain consistent. The D symmetry characteristics for both the ``force \revyxx{source}" \revy{quantities} and ``torque \revyxx{source}" \revy{quantities} are identical, with the D symmetry flipped in comparison to the fundamental $\text{X}=(\text{0})$ occurring in $\text{X}=(\text{da})$  and $\text{X}=(\text{pda})$, aligning with the D symmetry  broken criteria in particles. The \revchen{P symmetry} attributes of force and torque are inversed, yet their flipped characteristics concerning their respective $\text{X}=(\text{0})$ remain in accordance with the corresponding P-symmetry broken prerequisites in particles. Therefore, the P and D symmetry properties required for both particles and \revy{``force/torque source'' quantities} are exactly the same.}

In summary, we have built a universal and elegant framework \revlina{analyzing}
the mechanical effects \revlina{upon isotropic chiral particles of arbitrary sizes and compositions generated by generic 
monochromatic optical fields.  
\revlina{The light structure are completely characterized by 12 \revchen{``force/torque source" quantities} in quadratic form of $E$ and $H$ fields.
The quantities are further classified into 4 categories based on their \revchen{P and D symmetries} and
the effects beyond dipolar approximation are incorporated into 
the operators involving Laplacian, which apply to the 12 quantities and keep their P and D \revchen{symmetries.} 
The respective weights of the 12 quantities in optical force and torque are analytically derived and shown to depend exclusively on the particle properties. 
The rigorous and analytical \revchen{expressions} thus obtained for the optical force and torque are}
validated against the precise generalized-Mie theory results. 
The P and D symmetry-broken criteria are revealed
for the illuminated particle to couple with the \revchen{``force/torque source" quantities} to induce optical force/torque,
which offers a clear guideline for customizing optical manipulation techniques by designing appropriate structured light.} 
We hope our framework can inspire the development of a universal framework for non-reciprocal particles.
\begin{acknowledgments}
This work is supported by National Natural Science Foundation of China (No. 12204117 , No. 12174076 and No. 12074084); Guangxi Science and Technology Proiect (No. 2023GXNSFFA026002, No. 2024GXNSFBA010261, No. 2021GXNSFDA196001, and No. AD23026117); Open Project of State Key Laboratory of Surface Physics in Fudan University (KF2022\_15). X.L. is supported by the Research Grants Council of Hong Kong (16310422, AoE/P-502/20).
\end{acknowledgments}

\bibliography{ref.bib}

\end{document}
%